\documentclass[sigconf]{acmart}

\AtBeginDocument{%
  }

\usepackage{amsmath}
\usepackage{color}
\usepackage{url}
\usepackage{colortbl}
\usepackage{booktabs}
\usepackage{hhline}
\usepackage{multirow}
\usepackage{xspace}
\usepackage{hyperref}
\usepackage{cleveref}
\usepackage{booktabs}
\usepackage{longtable}
\usepackage{subfig}
\usepackage{float}
\usepackage{geometry}
\usepackage{tcolorbox}
\usepackage{tabularx}
\usepackage{pdflscape}  
\usepackage[table]{xcolor}
\usepackage{makecell}
\usepackage{enumitem}
\usepackage{svg}
\usepackage{arydshln}
\usepackage{graphicx}

\usepackage[colorinlistoftodos,textsize=footnotesize,textwidth=1.2cm]{todonotes}
\usepackage{marginnote}
\usepackage{paralist}

\newtoggle{withAppendix}
\newtoggle{withComments}
\toggletrue{withComments}
\usepackage{comment}

\makeatletter
\newcommand{\ifinsidefloat}[2]{\@ifundefined{@captype}{#2}{#1}}
\newrobustcmd{\annote@draw}[3]{{%
    \ifinsidefloat{%
    \marginnote{\todo[color=#3,inline]{#2: #1}}%
    }{%
    \todo[color=#3]{#2: #1}%
    }}}
\newrobustcmd{\annote}[3]{\iftoggle{withComments}{\annote@draw{#1}{#2}{#3}}{}}    
\makeatother

\newcolumntype{Z}{>{\hsize=.6\hsize\centering\arraybackslash}X}  
\newcolumntype{S}{>{\hsize=.4\hsize\centering\arraybackslash}X}  

\usepackage{tikz}
\newcommand{\highlightnum}[2]{%
  \protect\tikz[baseline]{%
    \node[
      rectangle,
      fill=#1!60,
      rounded corners=1pt,
      inner sep=1pt
    ] {\strut\normalfont #2};
  }%
}

\definecolor{mygreen0}{RGB}{235,245,238}
\definecolor{mygreen1}{RGB}{212,237,218}  
\definecolor{mygreen2}{RGB}{169,208,142}  
\definecolor{mygreen3}{RGB}{0,176,80}     

\definecolor{myred0}{RGB}{255,240,240}  
\definecolor{myred1}{RGB}{255,226,226}   
\definecolor{myred2}{RGB}{244,163,163}   
\definecolor{myred3}{RGB}{192,0,0}

\definecolor{lightgray}{rgb}{0.83, 0.83, 0.83}
\definecolor{darkgray}{rgb}{0.66, 0.66, 0.66}
\definecolor{lightergray}{rgb}{0.945,0.945,0.945}

\usepackage{siunitx}
\usepackage{xcolor}
\usepackage{colortbl}
\usepackage{booktabs}
\usepackage{multirow}

\copyrightyear{2026}
\acmYear{2026}
\setcopyright{cc}
\setcctype{by}
\acmConference[MSR '26]{23rd International Conference on Mining Software Repositories}{April 13--14, 2026}{Rio de Janeiro, Brazil}
\acmBooktitle{23rd International Conference on Mining Software Repositories (MSR '26), April 13--14, 2026, Rio de Janeiro, Brazil}
\acmDOI{10.1145/3793302.3793368}
\acmISBN{979-8-4007-2474-9/2026/04}

\begin{document}

\title{The Value of Effective Pull Request Description}
\author{Shirin Pirouzkhah}
\affiliation{%
	\institution{University of Zurich}
	\city{Zurich}
	\country{Switzerland}
}
\email{shirin@ifi.uzh.ch}

\author{Pavlína Wurzel Gonçalves}
\affiliation{%
	\institution{INESC TEC}
	\city{Porto}
	\country{Portugal}
}
\email{p.goncalves@ifi.uzh.ch}

\author{Alberto Bacchelli}
\affiliation{%
	\institution{University of Zurich}
	\city{Zurich}
	\country{Switzerland}
}
\email{bacchelli@ifi.uzh.ch}

\renewcommand{\shortauthors}{Pirouzkhah et al.}

\begin{abstract}
In the pull-based development model, code contributions are submitted as pull requests (PRs) to undergo reviews and approval by other developers with the goal of being merged into the code base.
A PR can be supported by a \emph{description}, whose role has not yet been systematically investigated.
To fill in this gap, we conducted a mixed-methods empirical study 
of PR descriptions. We conducted a gray literature review of guidelines on writing PR descriptions and derived a taxonomy of eight recommended elements. Using this taxonomy, we analyzed 80K GitHub PRs across 156 projects and five programming languages to assess associations between these elements and code review outcomes (e.g., merge decision, latency, first response time, review comments, and review iteration cycles). To complement these results, we surveyed 64 developers about the perceived importance of each element. Finally, we analyzed which submission-time factors predict whether PRs include a description and which elements they contain.
We found that developers view PR descriptions as important, but their elements matter differently: purpose and code explanations are valued by developers for preserving the rationale and history of changes, while stating the desired feedback type best predicts change acceptance and reviewer engagement. PR descriptions are also more common in mature projects and complex changes, suggesting they are written when most useful rather than as a formality.
\end{abstract}



\maketitle

\section{Introduction}
The pull-based development model is a popular method of collaborative development~\cite{ExploratoryOfPullBased}. This model supports the proposal and integration of code changes into a code base, and offers features for conducting a \emph{code review} of the proposed changes before they are merged.  Code review involves systematically examining code changes made by other developers to ensure quality, correctness, and adherence to coding standards~\cite{sadowski2018modern,shan2022using,wurzel2023competencies} through identifying bugs, improving code maintainability, and sharing knowledge among developers~\cite{bacchelliexpectations, mcintoshquality}. In the pull-based model, two roles stand out: the contributor (\textit{aka}, the PR author or requester) and the reviewer. The contributors submit code changes using PRs and can accompany their submission with a textual \emph{description}. The reviewers are tasked with evaluating these pull requests, possibly providing feedback to make further changes, and deciding whether they should be merged. A key enabler of this decision-making process is understanding the submitted code changes~\cite{bacchelliexpectations}. PR descriptions are an information source expected to facilitate this understanding process~\cite{comprehension}. However, large-scale evidence shows that PR descriptions are often left empty by developers~\cite{liuPRD}, thus raising questions on the actual role and usefulness of PR descriptions in the code review process in practice.

A wide range of sources---including industry and open-source community contribution guidelines, practitioner blogs and websites, technical reports, and white papers---seem to consider PR descriptions an important tool to support the review process. Many projects and organizations have established guidelines on how to write effective and complete pull request descriptions, in which they articulate best practices to ensure clarity, thoroughness, and relevance of code reviews~\cite{githubDocs, googledevelopers, atlassian}. For instance, according to GitHub Docs~\cite{githubDocs}, the goal of a PR description is to ensure that reviewers can quickly understand what the code change does. Google~\cite{googledevelopers} emphasizes that clear communication in PR descriptions is crucial, noting that a well-written summary increases the likelihood of faster acceptance. Atlassian~\cite{atlassian} highlights that detailed descriptions guide the reviewer through the code, saving time and making the review process faster. While these best practices are widely encouraged, the actual effect of PR descriptions on the effectiveness of the review process has not been empirically investigated. To address this gap, in this paper we present a mixed-methods empirical investigation into the value of PR descriptions. We first conducted a gray literature review (GLR) of guidelines and practitioner sources on writing PR descriptions, from which we derived a taxonomy of eight elements recommended for inclusion in PR descriptions. Building on this taxonomy, we conducted a large-scale analysis of 80K PRs in five programming languages across 156 projects on GitHub. We examined whether the presence of each PR description element shows statistically significant associations with code review outcomes, including merge decision, PR latency, first response time, number of review comments, and feedback-modification iterations. Observing that the effects of these elements varied across outcomes, we then surveyed 64 developers with diverse backgrounds in software development and code review to understand how practitioners perceive the importance of each PR description element. Finally, motivated by the findings of our data modeling and survey, we investigate which factors observable at submission time predict whether a PR includes a description and which description elements it contains. 

Our results show that developers generally perceive PR descriptions as important, yet individual elements prove more important in specific contexts. Descriptive elements such as purpose and code explanation are considered important by practitioners in most PRs, while interaction-oriented elements---such as specifying the type of feedback requested---are the most consistently predictive of positive review outcomes. In particular, indicating the feedback type increases the likelihood of a PR being merged and encourages richer reviewer engagement, despite being rated as important in fewer PRs than other elements. We further find that PR descriptions are more common in mature projects and complex code changes, suggesting that description writing is an adaptive, context-sensitive practice rather than a routine formality, supporting developers in the context of more formalized collaboration and more demanding code changes.

\vspace{-0.8em}
\section{Background}
\label{sec:background}
In this section we present background information and related work on pull-request descriptions and on the outcomes of the code review process, as they are both central to our investigation.

\subsection{PR Descriptions: Templates and Automation}
Contextual information to support the review of a proposed PR can be provided through multiple sources in addition to the PR description, such as issue trackers, documentation, and external websites. Past studies found that additional contextual information facilitates the review process, leading to quicker merge times and more positive outcomes in terms of code quality and reviewer satisfaction~\cite{confusion, comprehension}. Yet, when it comes to PR description, \citet{liuPRD} found that 34\% of PRs miss it. To tackle this problem, a growing body of research has focused on automatically generating PR descriptions~\cite{prhan, liuPRD, T5PRD, AIForDescription}. This line of work is motivated by concerns that missing descriptions deprive reviewers of important context~\cite{dougan}. However, there is limited evidence on how PR description content affects code review or what makes a PR description effective.

Beyond automation approaches, prior work has also examined approaches to improve PR descriptions with pull request templates. These templates were introduced by GitHub in 2016 as a mechanism to encourage contributors to provide structured information when opening a PR. A template is a predefined form stored in the repository that automatically appears in the PR description box, typically prompting authors to include items such as the purpose of the change, tests performed, and related issues. While templates are more frequently used in popular and active projects with large numbers of contributors, their adoption remains relatively low overall (about 1.2\% of GitHub repositories)~\cite{consistentTemplates}. The use of templates, however, has a positive impact on the project as it correlates with fewer duplicate PRs, shorter review times, and fewer review comments, especially when the templates are well-structured and emphasize essential elements like description and test information~\cite{consistentTemplates}.

\subsection{Code Review Process Outcomes}
\label{sec:outcomes}
We focus on six common outcomes that capture two dimensions of the code review process: Efficiency and interaction dynamics. The former is reflected by the merge decision, PR latency, and reopened PRs; the latter are captured by first response time, number of code review comments, and feedback.

\noindent\textbf{Merge decision:} Whether the proposed changes are ultimately integrated into the target code base is one of the most central outcomes in the pull-based development model. \citet{PRDecisionsExplained} found that the merge decision is primarily a matter of self-assessment--when contributors merge their own PRs. However, when contributors and reviewers differ, the strongest negative predictors of merge decisions include higher PR latency and the presence of review comments, whereas the strongest positive predictors include the number of commits and the existence of continuous integration in the PR. They also examined PR description length as a factor but found no statistically significant association with merge decisions.

\noindent\textbf{PR latency:}  
PR latency refers to the lifetime of a pull request from submission to closure. It is an important aspect in the code review process as it reflects the speed of software development~\cite{sadowski2019software}. Prior research has identified multiple factors influencing latency~\cite{LatencyExplained}, such as PR description length, change size, contributor experience, integrator availability, and the number of open PRs in the project. Among these factors, description length emerged as the most influential, explaining 46.3\% of the variance, indicating that more extensive textual descriptions are associated with longer review processes. On the other hand, \citet{wecannot} showed that accurately modeling review latency is difficult due to the inherent variability of the code review process. While these studies identify key factors shaping PR latency, they treat PR descriptions only in terms of length. Yet, the actual content of descriptions can be equally important, as specific elements can provide reviewers with the context needed to assess a change more efficiently.

\noindent\textbf{Reopened PRs:} 
Code review effectiveness can be likewise judged through the need to rework what has been done~\cite{oliveira2016software}. Reopened PRs usually result from unresolved issues with the code, such as bugs, change of minds, insufficient tests, test fails, or incompatible versions~\cite{jiang2019characteristics}. As one of the main purposes of code review is to prevent the introduction of defects in the code~\cite{bacchelliexpectations}, our study also evaluates PR reopening as another important marker of code review process effectiveness.

\noindent\textbf{First response time:}  
First response time refers to the interval between PR submission and the first reviewer comment. A faster response time is generally considered beneficial, shortens overall review latency, and signals reviewer availability. Response time is sensitive to several factors, including the presence of continuous integration~\cite{waitforit}, PR size~\cite{waitforit}, and submission timing~\cite{responselatency}. The content of PR descriptions may also influence how quickly a reviewer may respond: descriptions containing sufficient and well-structured contextual information may lower the cognitive effort required to understand a PR, thus reducing delays.

\noindent\textbf{Code review comments:}  
Code review comments are the primary medium of communication between authors and reviewers~\cite{referencingInDiscussions}. Through comments, reviewers request changes, seek clarification, and share concerns or opinions~\cite{codereviewQuestions}. Comments also contribute to improving documentation, maintainability, and overall code quality~\cite{VincentComments}. With experience, reviewers tend to provide more constructive and actionable feedback~\cite{usefulCRC, usefulnesscomments}. Both the absence and the abundance of comments can signal issues in the review process: few comments may indicate superficial evaluation~\cite{dougan}, while excessive discussion may reflect disagreement or coordination challenges~\cite{ wurzel2022interpersonal, egelman2020predicting}. Nevertheless, the number of review comments signifies how active the reviewer–author interaction is. Comprehensive and well-structured PR descriptions may provide the context necessary for focused, efficient, and meaningful feedback exchange.

\noindent\textbf{Feedback–modification iterations:}  
Feedback–modification iterations denote the number of times a PR is revised through additional commits made in response to reviewer feedback. PRs requiring revisions typically undergo multiple iterations; prior work shows that such PRs take longer to complete but are also more likely to be merged~\cite{modifyPRs}. \citet{bosu2015characteristics} define useful comments as those that trigger code changes, whereas excessive revision rounds can indicate inefficiency or coordination problems~\cite{dougan, wurzel2022interpersonal}.

\vspace{-0.8em}
\section{Research Questions}
Prior work has examined PR descriptions as part of the information available to reviewers during code review. However, much of this research has focused on surface characteristics of PR descriptions, such as their length or existence~\cite{PRDecisionsExplained,sadowski2018modern}. We aim to understand what constitutes an effective PR description and whether richer descriptions influence the code review process outcomes. Many practitioner sources and contribution guidelines emphasize that PR descriptions play an important role in the code review process, often recommending inclusion of specific elements to make descriptions clear, complete, and effective~\cite{googledevelopers, githubDocs, atlassian}.
However, these recommendations are fragmented. To address this, we ask the following research question:

\noindent\textbf{RQ$_1$: Which elements are recommended for inclusion in PR descriptions according to existing practitioner and community guidelines?}

We conducted a Gray Literature Review to collect what practitioners recommend to include in PR descriptions. However, it is unclear whether including such content is related to different code review process outcomes. Therefore, we ask:

\noindent\textbf{RQ$_2$: How does the presence of different PR description elements relate to code review outcomes?}

To answer this question, we performed a large-scale, data-driven analysis of 80K PRs from 156 GitHub projects using mixed-effects regression models to examine whether specific description elements correlate with review outcomes outlined in \Cref{sec:outcomes}. We also recognize that descriptions provide a context for reviewers' understanding, and therefore, we next investigate developers’ perceptions to understand how they value different aspects of PR descriptions and how their experiences align with existing guidelines. Building on this, we ask:

\noindent\textbf{RQ$_3$: How do developers perceive the importance and value of PR descriptions and their elements?}

To address RQ$_3$, we conducted an online survey with 64 developers to capture their perceptions of the importance of PR descriptions and their constituent elements during code review. The findings from RQ$_2$ and RQ$_3$ highlight that the importance of PR descriptions and their content is bound within a context. Building on these insights, we next explore what factors make contributors more or less likely to include descriptions and specific description elements in their PR submissions. This leads to our final two research questions:

\noindent\textbf{RQ$_4$: Which factors are associated with whether a PR includes a description?}

Answering RQ$_4$ helps to explain when authors provide any written context, which is a prerequisite for PR descriptions to support understanding.

\noindent\textbf{RQ$_5$: Which factors are associated with whether a PR description includes specific elements?}

RQ$_5$ goes one step further by explaining under which context each element is being included in the PR descriptions, helping to explain when authors provide different elements in their descriptions.

\vspace{-0.8em}
\section{Methodology}
We approached the listed research questions by performing a Gray Literature Review, a developer survey and a GitHub data analysis.
All the data we collected, and the materials we developed, are available in our replications package~\cite{reppackage}.

\subsection{Gray Literature Review (GLR)}  

To address RQ$_1$, we conduct a Gray Literature Review (GLR) following the guidelines of \citet{garousi2019guidelines}. We chose a GLR because community guidance on PR descriptions is largely practitioner-driven. Grayliterature includes non-peer-reviewed materials such as industry and community guidelines, documentation, practitioner blogs, and technical reports, which are commonly used to inform software engineering practice.

\noindent\textbf{Search strategy.}  
We systematically searched (i) general web search (Google); (ii) community sites (e.g., GitHub Docs, Atlassian, Google Developers); and (iii) snowballing from seed pages.  We used ``pull request description'', ``how to write pull request descriptions'', and ``best practices for PR descriptions'' as search terms. To ensure the reliability of results, we paid careful attention to selecting materials published by recognized companies, professional organizations, and experienced practitioners. To minimize the influence of personal search history or algorithmic bias, all searches were performed in Google’s incognito mode. A stopping criterion based on theoretical saturation was applied---when the last five retrieved sources introduced no new elements or practices.

\noindent\textbf{Selection and inclusion criteria.}  
We included sources that (i) explicitly provide guidance on writing PR descriptions, (ii) are publicly accessible, (iii) are aimed at software development practitioners, and (iv) present best practices in a form that can be broken down into distinct elements for analysis. Sources not directly related to PR descriptions (e.g., general code review advice without description-specific content) were excluded.

\noindent\textbf{Data extraction and synthesis.}  
From each selected source, the first author extracted the recommended practices and guidance items. We then applied open coding and iterative categorization to organize related practices into broader themes to arrive at a taxonomy of PR description elements, presented in the Results section (\autoref{tab:elements}).

\subsection{Software Repository Data Collection}

\begin{figure*}[t]
    \centering
    \caption{Overview of the data collection pipeline.}
    \includegraphics[width=\textwidth]{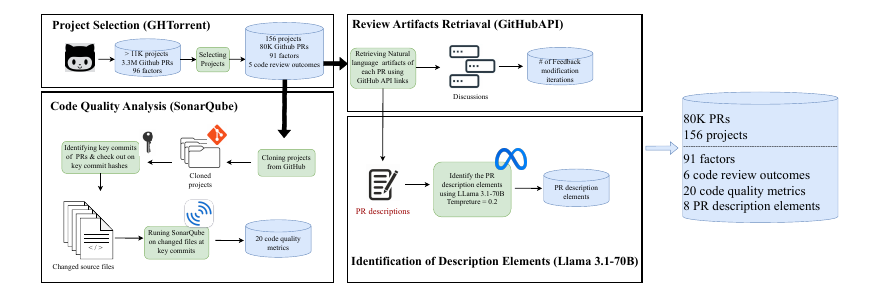}
    \label{fig:datapipeline}
\end{figure*}

Figure \ref{fig:datapipeline} illustrates the overall workflow of our data collection process. The pipeline integrates four main stages that collectively construct the dataset used in our analyses. We begin with the large-scale GHTorrent archive~\cite{PRDecisionsExplained}, from which projects and their PRs are selected. We rely on GHTorrent because it enables sampling of GitHub projects and pull requests across multiple repositories without the need to exhaustively crawl GitHub from scratch. While the latest publicly available GHTorrent release dates back to 2019, it still offers a large and diverse corpus of real-world PR activity. Furthermore, we enrich the data by extracting additional information about code quality (through SonarQube), textual review artifacts such as discussions and descriptions (through GitHub API), and description element identification using LLaMA 3.1-70B. Together, these stages yield a comprehensive dataset that combines information about PRs across 156 projects and 80K PRs.

\subsubsection{Project Selection (GHTorrent)}

To address $RQ2$, $RQ4$, and $RQ5$, we rely on the GHTorrent dataset that mirrors GitHub’s event stream. Specifically, we used the MySQL data dump dated June 1st, 2019, which has also been employed in prior work~\cite{PRDecisionsExplained}. This dataset captures pull requests and their associated metadata. It includes 96 factors relating to pull requests, the developers involved, and project-level characteristics, such as project age and size.
The dataset encompasses over 11K software projects and 3.3 million closed PRs, spanning diverse programming languages and project scales. We applied the following criteria to select projects from the GHTorrent dataset used in our analysis:

\noindent\textbf{Projects should have external contributors:} Projects dominated by core team members typically rely on shared knowledge, resulting in less detailed PR descriptions. To capture a more representative set of PR descriptions, we included only projects where at least one-third (33\%) of PRs were submitted by external contributors.
    
\noindent\textbf{Projects should include tests:} One of the key description elements recommended in the guidelines identified in the GLR is how the changes were tested. Therefore, we required projects that contain test files. This requirement applies at the project level: individual PRs may or may not include tests, but they must belong to projects that do. 

\noindent\textbf{Projects should not be smaller than 200 PRs:} To avoid toy projects and ensure substantial activity, we require projects to have at least 200 PRs in the dataset.

\noindent\textbf{PRs should have at least one reviewer:}  PRs must be reviewed by someone other than the contributor. Because PR descriptions are most important when facilitating understanding for others, we excluded self-reviewed PRs.

\noindent\textbf{PRs should modify at least one source code file:} We excluded PRs that only changed non-code artifacts.

\smallskip 

These criteria yielded a dataset of 156 projects and 80K PRs.

\subsubsection{Code Quality Analysis (SonarQube)}
PR descriptions are intended to help reviewers understand and navigate the review of a code change. To control for technical factors that may independently influence review outcomes, we collected static code quality and complexity metrics. 
Aspects such as the size of modifications, the presence of code quality issues, or the structural complexity of the changes can all influence the amount of additional explanation reviewers require.
Since the GHTorrent dataset does not include the source code files needed to compute these metrics, we extended our pipeline beyond GHTorrent. Starting from the 156 projects selected from GHTorrent, the second stage retrieves the corresponding source code and computes quality and structural properties of the PR code changes. Using the repository links of the selected projects, we cloned these repositories to access their full version histories. For each pull request, we identified key commits and used their hashes to check out the state of each changed source code file at the time of PR submission. This ensured that the analyzed files reflected the precise code changes proposed in each PR at the submission time. Once the relevant files were checked out, we subjected them to static analysis using SonarQube, a widely used static analysis platform, accessed via the Sonar Scanner tool~\cite{SonarQube}. SonarQube produced detailed reports capturing a range of file-level code quality metrics. For Java projects, Sonar Scanner typically requires a successful project build. However, due to configuration and compatibility issues, it was not always feasible to build every cloned repository. Since our focus was on file-level metrics, we adapted our process by analyzing each modified file in each PR independently rather than relying on a full project build. For PRs involving multiple modified files, we aggregated file-level metrics to compute a cumulative value representing the entire PR. This step resulted in 20 code quality metrics related to each PR.

\subsubsection{Review Artifacts Retrieval (GitHub API)}

The third stage of the pipeline collects the natural language artifacts produced in the review process, namely the PR descriptions, commit messages, review comments, and discussion threads through the GitHub API. These artifacts provide the conversational and descriptive context surrounding each PR. GitHub exposes review interactions through two main comment types: (1) inline review comments attached to specific lines in a diff and (2) overall review comments posted in the PR conversation thread. Although these comment types differ in where they appear and what they reference, both represent reviewer--author communication and contribute to review interaction. Since our goal is to capture overall review engagement, we did not separate these two categories in our analysis. Instead, we aggregated them and treated the total number of comments exchanged during the PR as a proxy for reviewer--author interaction intensity.

To reconstruct the event history of each PR, we used all review comments and all commits pushed to the PR branch. We then aligned these events into a single chronological timeline using their timestamps, producing an ordered sequence that captures how the review unfolded over time (e.g., PR opened → reviewer comment → author commit → reviewer comment → \ldots). Review comments are typically human-written, but they can also originate from bots. To avoid mischaracterizing such automated output as human feedback, we explicitly accounted for bot activity in these review comments. We heuristically identified bot-generated review interactions by matching the usernames against a curated list of common GitHub bot accounts and naming patterns. We represent this reconstruction as a single timestamp-ordered sequence of events
\(\mathcal{E} = \langle e_1, e_2, \dots, e_n\rangle\),
where each event \(e_i\) is either a human comment or a commit. From this reconstruction, we derived the number of ``Feedback–modification iterations'' as cycles in which  review comments are followed by at least one subsequent revision, observable as new or altered commits submitted in direct response.

\subsubsection{Identification of PR Description Elements (LLaMA 3.1–70B;)}

In the final stage, we used the retrieved PR descriptions as input to the LLaMA 3.1-70B model to automatically identify which elements from our taxonomy were present in each description. This automated identification and classification allowed us to extract the elements defined in the GLR at scale.
We framed the task as a simple binary classification (present / not present) for each element. To ensure consistency and reduce randomness in the outputs, all prompts were executed with a temperature of 0.2, as recommended for extraction or classification tasks~\cite{LLMTempreture}. To ensure the reliability of the LLM-based extraction, we manually validated the model’s outputs. A fully random sample of 100 PR descriptions was annotated by the first author using the taxonomy of PR description elements. Considering the formula for sample size estimation~\cite{samplesad}, this gives us a confidence level of 95\% that the real ability of the LLM in identifying PR description elements is within $\pm$10\% of what we measured. Cases that were ambiguous or difficult to classify were discussed with the second author, after which we produced a single adjudicated (``gold'') human annotation for all 100 descriptions. We then compared this gold annotation to the classifications produced by LLaMA 3.1–70B, obtaining an average precision of 0.85 across elements (lowest for ``reason of the PR'' element, precision = 0.30). In our taxonomy, ``reason of the PR'' denotes why the PR is being made. Identifying reason often requires inferring intent—capabilities that LLMs are known to struggle with in extraction settings. Because the model’s reliability for reason was low, we excluded this element from subsequent analyses.

As summarized in Figure \ref{fig:datapipeline}, our data collection pipeline extends the information available in GHTorrent archive by adding several new sources of data. Beyond the original 96 factors provided by GHTorrent, we contributed for each PR 20 code-quality metrics from SonarQube analyses, natural-language artifacts (discussions, commit messages, review comments, and PR descriptions), the number of review iterations, and LLaMA-based binary labels indicating the presence or absence of each element in the PR description.

\subsection{Mixed-Effects Regression Models}
\label{sec:regression_models}

To examine the role of PR descriptions and their content in the context of projects and the code review process, we employed a hierarchical mixed-effects regression modeling approach. We built several models to identify potential intervening variables and answer our Research Questions. We summarize the steps of the analysis below:

\noindent\textbf{Collinearity analysis:} Prior to running any of the regression models, we examined correlations among all continuous and categorical factors to identify potential collinearity. Specifically, we computed the Spearman correlation coefficient (\(r\)) for continuous factors, Cramér's V (\(F_c\)) for categorical factors, and partial Eta-squared (\(\eta^2\)) for mixed continuous–categorical relationships. We considered \(r > 0.7\), \(F_c > 0.5\), and \(\eta^2 > 0.14\) as indicators of strong correlation and removed such factors to prevent redundancy. Additionally, we evaluated multicollinearity using the Variance Inflation Factor (VIF), excluding predictors with \( \text{VIF} \geq 5 \) to avoid inflated variance and unstable estimates.

\noindent\textbf{Regression Modelling:} We applied the regression models based on the outcome type. We used mixed-effects regression models, implemented with the \textit{lmer} and \textit{glmer} functions from the \textit{lme4} package in R, depending on whether the dependent variable was continuous (e.g., PR latency) or binary (e.g., merge decision). Mixed-effects regression was selected to account for the hierarchical structure of the data, in which pull requests are nested within projects. The project identifier was included as a random effect to model intra-project dependencies, capturing shared variance among pull requests within the same project.

\noindent\textit{Outcome Model:} First, we identify which variables explain the variation in code review outcomes (e.g., merge decision) to identify control variables for further analysis. Each model included a code review outcome as the dependent variable, with all available project-, developer-, PR-, and code-level factors as fixed effects. Only factors that passed the collinearity and multicollinearity checks were retained in the models. PR description elements were not included in this step. The project identifier was included as a random effect. The goal of this step was to determine which variables significantly influence code review outcomes. For every outcome, the top three factors with the largest variance contribution in explaining that outcome are identified as \textbf{outcome control variables}.

\noindent\textit{Element Model:} This model examined which factors, observable at the time of PR submission, predict whether a PR description includes specific elements from our taxonomy (e.g., purpose), thus answering $RQ5$. Only factors that passed the collinearity and multicollinearity checks were retained in the models. For each element, we fitted a binary mixed-effects regression with the presence or absence of that element as the dependent variable and pre-submission project-, developer-, PR-, and code-level variables as fixed effects. The project identifier was included as a random effect. This step identified variables that explain the likelihood of element inclusion in the PR description. We identified \textbf{global control variables} as the subset of most consistent positive and negative predictors that appeared repeatedly as significant across all Element Models (Table~\ref{tab:summary_elements_predictors_singlecol}). These common predictors are: \textit{diff size}, \textit{test inclusion in PR}, \textit{number of commits}, \textit{number of contributors’ previous PRs}, \textit{contributors’ daily workload}, and \textit{number of code smells}. 

\noindent\textit{Controlled Model:} This model is the main explanatory model and tests the relationship between PR description elements and code review outcomes while controlling for confounding factors. For each outcome, we combined \textbf{outcome control variables} specific to that outcome and \textbf{global control variables} for PR description elements to construct a comprehensive set of control variables.

\noindent\textit{Baseline Model:} We also built baseline models. The baseline model includes only the six PR description elements as fixed effects and the project identifier as the random effect, without any additional controls. This model served as a reference to assess the direct, unadjusted association between description elements and review outcomes, providing a benchmark for evaluating the incremental explanatory power of the controlled model.

\subsection{Developer Survey}
We designed and conducted an online survey that aimed to gather insights into developers’ perceptions of PR descriptions and the importance they assign to each description element (\Cref{tab:elements}). The survey began with an introduction and data handling information, followed by questions about the participants' professional backgrounds and the frequency and quality of PR descriptions they encountered. In the third step, respondents were asked to define what they consider a ``good'' PR description. This prompt encouraged participants to reflect and articulate their own mental model of a good description, which then served as a reference point for the subsequent questions. In the fourth step, participants reflected on their past review experiences by answering: ``For the PRs you reviewed, how important was the presence of a description?'' They also explained the rationale behind their ratings. In the fifth step, participants evaluated the importance of key elements of PR descriptions from our taxonomy (\autoref{tab:elements}). They were asked: ``In how many PRs were the following elements in the descriptions important to you?'' This phrasing was chosen to account for the fact that not every PR requires every element. By asking about frequency across their experience, rather than assigning a single importance rating, we captured a more realistic view of how often each element matters in practice.
To reach a diverse pool of respondents, we distributed the survey across multiple software development communities, including Reddit and LinkedIn groups, targeting developers with varying backgrounds and levels of experience.

\noindent\textbf{Survey Pilot Runs:}
Before publicly releasing the survey, we conducted a pilot run to validate the survey's clarity and reliability. The pilot study involved seven participants, who were recruited through the professional networks of the study authors. We updated the survey according to their feedback by refining the wording of questions and adding explanatory text to ensure participants understood the context of the questions.

\noindent\textbf{Survey Analysis}
We analyzed the survey responses by validating the raw data and removing duplicate submissions, marking incomplete responses (e.g., those missing all closed questions), and checking for inconsistent patterns (for example, respondents who selected ``never'' for review frequency but mentioned reviewing PRs ``once a week''). For the closed-ended questions (such as the importance of PR description elements), we calculated the percentage of respondents selecting each option rather than reporting raw counts. For the open-ended question of the definition of good PR description, we conducted a close coding approach. Each definition was examined line by line and mapped to the elements in our PR description taxonomy.

\vspace{-0.8em}
\section{Results}
\textbf{RQ$_1$: Which elements are recommended for inclusion in PR descriptions according to existing practitioner and community guidelines?}
To answer RQ$_1$, we gathered practitioner recommendations through a GLR. We retrieved 22 relevant sources and derived a taxonomy of eight elements that are recommended to be included as best practices for writing PR descriptions. \autoref{tab:elements} summarizes these elements, providing their definitions and the sources in which they were recommended. The final set of description elements ranged from stating the purpose of the change and explaining the implementation to requesting a specific type of feedback from the reviewers. The ``Screenshots'' element was recommended as an optional element for PRs that modify the user interface (UI), front-end layout, or visual presentation. As this element is context-dependent, we excluded it from the subsequent analyses of PR data. This taxonomy serves as the basis for designing the developer survey and for the repository data analysis.

{\footnotesize
\begin{table*}[h!]
  \centering
  \caption{Taxonomy of PR Description Elements Derived from Gray Literature Review}
  \label{tab:elements}
  \begin{tabular}{p{4.5cm}p{6.5cm}p{5cm}}
\hline
\textbf{Element} & \textbf{Definition} & \textbf{Reference}  \\ \hline
Purpose of the PR (Purpose) & Explains the goal of the changes & \cite{Gonzalo, Pavel, Julio, Sajal, Araújo, Mercedes, James, Kara, Codacy, Michael, nimblehq, Mark, Bomberbot, atlassian, googledevelopers, githubDocs} \\
Reason of the PR (Reason) & Explains why the change is necessary & \cite{Gonzalo, Pavel, Julio, Sajal, Alastair, Araújo, Mercedes, James, Kara, Arthur, Erik, Michael, vanillaforums, Mark, Jay, atlassian, githubDocs} \\
Explanation of the Code Changes (Cod Exp) & Explains details of changed code & \cite{Gonzalo, Pavel, Julio, Sajal, Alastair, Araújo, Mercedes, James, Arthur, GitLab, Erik, Codacy, vanillaforums, nimblehq, Mark, Jay, Bomberbot, atlassian, googledevelopers, githubDocs} \\
Link to Related Issue(s) (Link) & Provides traceability by linking to relevant issues & \cite{Pavel, Julio, Sajal, Mercedes, James, Arthur, Erik, nimblehq, Mark, Bomberbot, atlassian, googledevelopers, githubDocs} \\
Type of Feedback Needed (Feedback Type) & Specifies the type of feedback the author expects from reviewers & \cite{Julio, Araújo, githubDocs, googledevelopers}  \\
Guidance on the Order of File Reviews (Order) & Suggests the order in which changed files should be reviewed & \cite{Julio, googledevelopers}  \\
Explanation of Tests (Test Exp) & Explains how changes were tested or refers to relevant tests & \cite{Gonzalo, Julio, Sajal, Alastair, Mercedes, Kara, Arthur, Codacy, Michael, nimblehq, Jay, Bomberbot, githubDocs} \\
Screenshots & Provides visual evidence & \cite{Gonzalo, Julio, Alastair, Araújo, Mercedes, James, Arthur, GitLab, Erik, Codacy, nimblehq, Mark, Bomberbot} \\ \hline
\end{tabular}
\end{table*}
}

\noindent\textbf{RQ$_2$: How does the presence of different PR description elements relate to code review outcomes?}

Observing the 80K PRs, we found that PR descriptions commonly include the elements providing descriptive information, such as the purpose of the code change (54.54\% of PRs), code explanations (55.39\%) and a link to the issue (45.16\%) and only a small portion of the PRs containing tests also mention them in the description (13.66\%). The elements focusing on directing reviewers in their code review, such as requesting a specific kind of feedback (16.20\%) and suggesting the order in which to review the files (6.10\%), were used in a smaller portion in PR descriptions. 

To investigate whether the content of PR descriptions can be linked to certain code review outcomes, we used mixed-effects regression, while controlling for potential factors. Table~\ref{tab:pr_outcomes_with_metric} presents the results of both the \emph{baseline} and \emph{controlled} models for predicting each outcome, and the \textit{Outcome Controls} column presents outcome control variables based on the \emph{outcome} model described in Section~\ref{sec:regression_models}. For continuous outcomes, we report linear mixed-effects models results, where the \textit{Coeff.} column displays the estimated regression coefficients, indicating the direction and magnitude of each predictor’s effect on the outcome—larger absolute values correspond to a stronger influence. For the binary outcome (\textit{Merge Decision}), we employ a logistic mixed-effects model, and the effect size of each predictor is also expressed as an \textit{odds ratio (OR)}, computed as the exponential of its estimated coefficient ($\mathrm{OR}_j = \exp(\hat{\beta}_j)$), using the Wald method~\cite{wald}. An OR greater than~1 implies that higher values of the predictor increase the likelihood of a PR being merged, while an OR below~1 indicates the opposite effect. Furthermore, because the \textit{Order} element is only meaningful when a PR modifies multiple files, and the \textit{Test} element is relevant only when tests are included, we incorporated interaction terms linking these elements to their respective contextual conditions (\textit{more than one file changed} and \textit{test inclusion indicator}). This specification ensures that the effects of these elements are evaluated only in contexts where they are conceptually applicable, thereby preventing spurious associations that could arise from PRs where these elements have no functional relevance. Each element is highlighted in green only when it reached statistical significance in both the baseline and controlled models for a given outcome.

All the PR description elements were extracted from practitioner guidelines on how to provide valuable context for reviewers. However, when we observed the relationship of these elements with various code review outcomes, we found only limited evidence for their relationship. While many of the elements showed significant relationships with code review outcomes, the effect sizes were generally negligible with several exceptions. First, PRs where the description contains \textit{Code Explanation} are approximately 12–20\% more likely to be merged compared to those without such information, suggesting code explanations increase merge likelihood. The \textit{Feedback Type} element—despite being present in only 16.20\% of PR descriptions—shows significance across all outcomes except reopened PRs. Based on the estimated coefficients, its influence varies in strength across outcomes. The estimated coefficients for the number of review comments and PR latency are moderately large, suggesting that PRs whose descriptions ask for specific feedback take slightly longer to be closed and also contain longer discussions, reflecting increased reviewer engagement. Interestingly, the estimated coefficients and odds ratios indicate that PRs where the author explicitly requests or frames the needed feedback are 64–72\% more likely to be merged compared to those without such cues.

\begin{table*}[t]
\centering
\small
\captionsetup{justification=centering}
\caption{Comparison of baseline and controlled models for PR description element effects on code review outcomes.
Each cell shows the estimated coefficient and corresponding significance level based on Holm-corrected \emph{p}-values.}
\label{tab:pr_outcomes_with_metric}
\resizebox{\textwidth}{!}{
\begin{tabular}{p{1.5cm}|p{3cm}|l l|c|c|c|c|c|c|c}
\hline
\textbf{Outcome} &  \textbf{Outcome Controls} & \textbf{Model} &  & \textbf{Intercept} & \textbf{Purpose} & \textbf{Code Exp.} & \textbf{Link} & \textbf{Feedback Type} & \textbf{Order} & \textbf{Test Exp.} \\ 
\hline
 Presence(\%)& - & - & - & - & 
\textbf{54.54\%} & 
\textbf{55.39\%} & 
\textbf{45.16\%} & 
\textbf{16.20\%} & 
\textbf{6.10\%} & 
\textbf{13.66\%} \\ 
\hline

\multirow{4}{*}[-1.2ex]{\makecell[l]{First \\ Response \\ Time}}  & \multirow{4}{*}[-1.2ex]{\makecell[l]{1-\#of open PRs in project\\ 2-Reviewer’s + emotion\\ 3-CI exists}} 
 & \multirow{2}{*}{Baseline}   & Coeff.    & 0.509 & \highlightnum{mygreen2}{0.068} & 0.000 & 0.021 & \highlightnum{mygreen2}{0.082} & 0.430 & 0.088 \\[-0.3ex]
 & &                                       & Sig. & ***   & \highlightnum{mygreen2}{***}   &       &      & \highlightnum{mygreen2}{***}   &       & ** \\ 
\cdashline{3-11}
 & & \multirow{2}{*}{Controlled} & Coeff.  & 0.508 & \highlightnum{mygreen2}{0.035} & 0.004 & 0.004 & \highlightnum{mygreen2}{0.068} & 0.359 & 0.057 \\[-0.3ex]
 & &                                      & Sig. & *** & \highlightnum{mygreen2}{***} &  &  & \highlightnum{mygreen2}{***} &  &   \\ 
\hline

\multirow{4}{*}[-1.2ex]{\makecell[l]{2- Feedback \\ Modification \\ Iterations}} & \multirow{4}{*}[-1.2ex]{\makecell[l]{1-\# of participants \\ 2-\# of commits\\ 3-Core memeber comment}} 
 & \multirow{2}{*}{Baseline}   & Coeff.    & 0.392 & \highlightnum{mygreen2}{0.109} & 0.005 & \highlightnum{mygreen2}{0.059} & \highlightnum{mygreen2}{0.114} & 0.240 & \highlightnum{mygreen2}{0.174}  \\[-0.3ex]
 & &                              & Sig.    & *** & \highlightnum{mygreen2}{***}   &    &   \highlightnum{mygreen2}{***}    &  \highlightnum{mygreen2}{***}    &     &   \highlightnum{mygreen2}{***}     \\ 
\cdashline{3-11}
 & & \multirow{2}{*}{Controlled} & Coeff.    &-0.117 & \highlightnum{mygreen2}{0.047} & 0.008 & \highlightnum{mygreen2}{0.047} & \highlightnum{mygreen2}{0.067} & 0.063 & \highlightnum{mygreen2}{0.011}  \\[-0.3ex]
 & &                                       & Sig.    & *** & \highlightnum{mygreen2}{***}   &    &   \highlightnum{mygreen2}{***}    &  \highlightnum{mygreen2}{***}    &     &   \highlightnum{mygreen2}{***}     \\  
\hline

\multirow{4}{*}[-1.2ex]{\makecell[l]{3- Number \\ of Review \\ Comments}} & \multirow{4}{*}[-1.2ex]{\makecell[l]{1-\# of participants\\ 2-Core member comment\\ 3-\# of contrib's\\ previous PRs }} 
 & \multirow{2}{*}{Baseline}   & Coeff.    & 0.321 & \highlightnum{mygreen2}{0.133} & -0.016 & \highlightnum{mygreen2}{0.051} & \highlightnum{mygreen2}{0.210} & 0.306 & \highlightnum{mygreen2}{0.148}  \\[-0.3ex]
 & &                               & Sig.    & *** & \highlightnum{mygreen2}{***}   &    &   \highlightnum{mygreen2}{***}    &  \highlightnum{mygreen2}{***}    &     &   \highlightnum{mygreen2}{***}     \\  
\cdashline{3-11}
 & & \multirow{2}{*}{Controlled} & Coeff.    & -0.504 & \highlightnum{mygreen2}{0.051} & 0.000 & \highlightnum{mygreen2}{0.046} & \highlightnum{mygreen2}{0.145} & 0.085 & \highlightnum{mygreen2}{0.075} \\[-0.3ex]
 & &                               & Sig.    & *** & \highlightnum{mygreen2}{***}   &    &   \highlightnum{mygreen2}{***}    &  \highlightnum{mygreen2}{***}    &     &   \highlightnum{mygreen2}{***}     \\  
\hline

\multirow{4}{*}[-1.2ex]{\makecell[l]{4- PR \\ Latency}} & \multirow{4}{*}[-1.2ex]{\makecell[l]{1-Conflict in comments\\ 2-\#of open PRs in project\\3-Core memeber comment }} 
 & \multirow{2}{*}{Baseline}   & Coeff.    & -0.013 & \highlightnum{mygreen2}{0.138} & \highlightnum{mygreen2}{0.035} & \highlightnum{mygreen2}{0.077} & \highlightnum{mygreen2}{0.262} & 0.289 & 0.111  \\[-0.3ex]
 & &                               & Sig.    &  & \highlightnum{mygreen2}{***} & \highlightnum{mygreen2}{***} & \highlightnum{mygreen2}{***} & \highlightnum{mygreen2}{***} &  & **     \\ 
\cdashline{3-11}
 & & \multirow{2}{*}{Controlled} & Coeff.    & -0.268 & \highlightnum{mygreen2}{0.072} & \highlightnum{mygreen2}{0.027} & \highlightnum{mygreen2}{0.047} & \highlightnum{mygreen2}{0.176} & 0.202 & 0.042 \\[-0.3ex]
 & &                                 & Sig.    &  & \highlightnum{mygreen2}{***} & \highlightnum{mygreen2}{***} & \highlightnum{mygreen2}{***} & \highlightnum{mygreen2}{***} &  &   \\ 
\hline

\multirow{4}{*}[-1.2ex]{\makecell[l]{5- Merge \\ Decision}} & \multirow{4}{*}[-1.2ex]{\makecell[l]{1-Contri\&reviewer \\prior interaction\\ 2-Project age\\ 3-PR is from core member}} 
 & \multirow{3}{*}{Baseline}   & Coeff.    & 0.243 & -0.022 & \highlightnum{mygreen2}{0.116} & 0.226 & \highlightnum{mygreen2}{0.439} & -0.556 & 0.069  \\[-0.3ex]
 & &                             & OR & 1.276 & 0.978 & \highlightnum{mygreen2}{1.123} & 1.253 & \highlightnum{mygreen2}{1.645} & 0.573 & 1.072 \\[-0.3ex]
 & &                             & Sig.    & ** &  & \highlightnum{mygreen2}{*} & *** & \highlightnum{mygreen2}{***} &  &     \\ 
\cdashline{3-11}
 & & \multirow{3}{*}{Controlled} & Coeff.    & 1.472 & 0.010 & \highlightnum{mygreen2}{0.187} & 0.087 &  \highlightnum{mygreen2}{0.326} & -1.046 & 0.009 \\[-0.3ex]
 & &                              & OR.    & 4.362 & 1.010 & \highlightnum{mygreen2}{1.206} & 1.091 & \highlightnum{mygreen2}{1.721} & 0.351 & 1.009\\[-0.3ex]
 & &                              & Sig.    & *** &  & \highlightnum{mygreen2}{***} &  & \highlightnum{mygreen2}{***} &  &    \\ 
\hline

\multirow{4}{*}[-1.2ex]{\makecell[l]{6- Reopened \\ PR}}  & \multirow{4}{*}[-1.2ex]{\makecell[l]{1-Length of discussion\\ 2-\# of commits\\ 3-Contrib's + emotion  }} 
 & \multirow{3}{*}{Baseline}   & Coeff.    & -0.385 & 0.133 & -0.068 & 0.046 & -0.087 & -0.616 & -0.001  \\[-0.3ex]
 & &                             & OR & 0.681 & 1.143 & 0.934 & 1.047 & 0.917 & 0.540 & 0.999  \\[-0.3ex] 
 & &                               & Sig.    & ** &  &  &  &  &  &     \\ 
\cdashline{3-11}
 & & \multirow{2}{*}{Controlled}           & Coeff.    & -1.425 & 0.0713 & -0.142 & -0.042 & -0.171 & -1.185 & -0.079 \\[-0.3ex]
 & &                                       & OR     & 0.240 & 1.07 & 0.867 & 0.958 & 0.842 & 0.305 & 0.923 \\[-0.3ex]
 & &                                       & Sig. & *** &  &  &  &  &  &   \\ 
\hline

\end{tabular}
}
\vspace{0.5em}
\scriptsize
\textbf{1- } \textbf{[Baseline:} $R^2_m = 0.003$ \quad $R^2_c = 0.154$ \quad  \textbf{Controlled:} $R^2_m = 0.027$ \quad $R^2_c = 0.170$] \quad
\textbf{2- }  \textbf{[Baseline:} $R^2_m = 0.017$ \quad $R^2_c = 0.140$ \quad \textbf{Controlled:}$R^2_m = 0.131$ \quad $R^2_c = 0.220$] \\
\textbf{3- }  \textbf{[Baseline:} $R^2_m = 0.015$ \quad $R^2_c = 0.179$ \quad \textbf{Controlled:} $R^2_m = 0.179$ \quad $R^2_c = 0.303$] \quad
\textbf{4- }  \textbf{[Baseline:} $R^2_m = 0.015$ \quad $R^2_c = 0.221$ \quad \textbf{Controlled:} $R^2_m = 0.182$ \quad $R^2_c = 0.322$]\\
\textbf{5- }  \textbf{[Baseline:} $R^2_m = 0.006$ \quad $R^2_c = 0.234$ \quad \textbf{Controlled:} $R^2_m = 0.259$ \quad $R^2_c = 0.464$] \quad
\textbf{6- }  \textbf{[Baseline:} $R^2_m = 0.001$ \quad $R^2_c = 0.170$ \quad \textbf{Controlled:} $R^2_m = 0.148$ \quad $R^2_c = 0.268$]
\end{table*}

\textbf{RQ$_3$: How do developers perceive the importance and value of PR descriptions and their elements?}

Despite these elements being broadly recommended in the practitioner community, we could not observe many meaningful relationships between such information in PR descriptions with code review outcomes. Furthermore, we have observed a disparity in how many PRs include different description elements. Therefore, we wanted to investigate whether these elements are actually important to developers and what value they provide to them.

To answer RQ$_3$, we designed an online developer survey to capture both developers’ general perceptions of PR descriptions and their views on how often the specific elements from our taxonomy are important in practice. We collected a total of 64 survey responses, out of which 57 were valid. 
To contextualize our survey findings and assess their representativeness, we compared key demographics of the survey sample with the population reflected in our dataset from GHTorrent. Among the survey respondents, 81\% were based in Europe, whereas the dataset from GHTorrent is more geographically diverse, with 14\% of contributors from Europe and 39\% from the United States (the remainder distributed across other regions). The survey sample was also male-dominated (female-to-male ratio 11:89), which is broadly comparable to the gender distribution in the dataset from GHTorrent (7:93). In contrast, the distribution of respondents’ professional roles closely matches the role distribution observed in our dataset. Moreover, among the survey respondents, 78\% reported having more than three years of software development experience. Most respondents (81\%) perform code review at least once a week, and (75\%) submit their code changes for review at least once a week. 54\% reported that the majority of the PRs they received for review included descriptions, and half of these descriptions were of good quality. We observed that PR descriptions are truly important to developers. 60\% of survey respondents reported PR descriptions as very or extremely important, 19\% as moderately important, and 19\% as slightly important, while only 3\% considered them not at all important. Developers highlighted four key reasons for PR description importance:

\noindent\emph{Understanding:} Clear descriptions provide essential background for reviewers and reduce misunderstandings. They explain bug fixes, implementation details, and design decisions. 

\noindent\emph{Historical tracking:} They serve as documentation for future reference and audits.

\noindent\emph{Code navigation:} They help reviewers locate and focus on relevant parts in complex code changes.

\noindent\emph{Review Efficiency:} Through the other reasons, reviewers can reach better reviewing efficiency.
    
\begin{figure}[t]
    \centering
    \caption{Percentage of PRs in which developers perceive a given description element as important.}
    \includegraphics[width=1\linewidth]{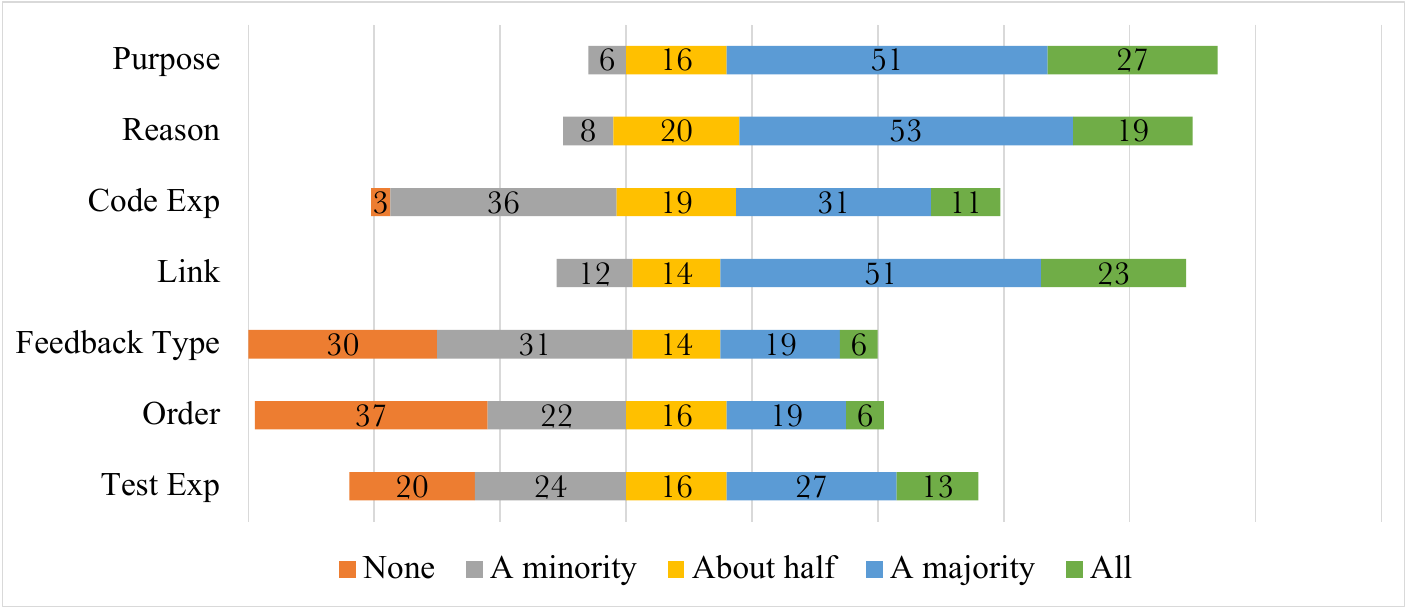}
    \label{fig:descriptives2}
\end{figure}

\noindent The survey results corroborate the observation from repository data analysis---the PR description elements are not always important to include. \Cref{fig:descriptives2} shows that \textit{purpose of the PR}, \textit{reason of the PR}, and \textit{links to related issues} are identified as consistently valuable elements in most or all of the PRs developers encountered. Elements such as explanation of code changes or how changes were tested are moderately valued. In contrast, elements like \textit{type of feedback needed} and \textit{guidance on the order of file reviews} were least frequently considered important. Overall, these results suggest that the PR descriptions elements descriptive of the \emph{code changes} are universally important while the elements focused on the reviewer interaction and direction are considered important in a smaller portion of PRs.


\textbf{RQ$_{4}$: Which factors are associated with whether a PR includes a description?}

The results of RQ$_2$ and RQ$_3$ suggest that PR descriptions and the specific elements are important only in certain contexts. Therefore, we aimed to understand which factors are related to the presence of PR descriptions in the first place and to the inclusion of specific elements (addressed in RQ$_5$).

\autoref{tab:has_description} summarizes the results of the mixed-effects logistic regression analysis examining which factors observable at the PR submission time are significantly associated with the likelihood of a PR including a description.
The \textit{Effect Size} column reports the estimated \textit{odds ratio (OR)} for each predictor. The \textit{95\% CI} column provides the upper and lower bounds of the 95\% confidence interval around the estimated odds ratio. The table is sorted by the magnitude of the effect size, with the most positive and most negative predictors highlighted in green and red, respectively. Model fit statistics show that although fixed effects alone explain a limited proportion of the variance ($R^2_m = 0.027$), the inclusion of project-level random effects substantially improves model performance and discriminative power($R^2_c = 0.246$). This indicates that the effect of the fixed predictors is contextualized by project-specific factors—that is, their effects vary across projects depending on characteristics such as documentation culture, review norms, and team practices. In other words, the predictors identified as significant are not universally influential, but rather project-dependent in how they shape the likelihood of including a PR description.

PRs that belong to more mature projects (project age at PR submission time), contain code of higher cognitive complexity, and include tests are more likely to contain a written description. Likewise, PRs from contributors who express more positive emotion tend to feature better-documented PRs. In contrast, PRs that delete more files and contain more code smells are less likely to include a PR description. Interestingly, PRs submitted by more experienced contributors—those who have previously submitted many PRs—and PRs from projects with higher overall success rates (i.e., a greater proportion of merged PRs) are also less likely to include descriptions. This pattern may reflect a degree of informal efficiency: in well-established or highly successful projects, and among experienced contributors, shared understanding and trust may reduce the perceived need for explicit documentation, leading to more succinct or omitted PR descriptions.

\begin{table}[htbp]
\centering
\scriptsize
\caption{Regression results for the effect of pre-submission factors on the likelihood of including any description in PRs.}
\label{tab:has_description}
\resizebox{0.48\textwidth}{!}{%
\begin{tabular}{lccr}
\hline
\textbf{Variable} & \textbf{Effect Size} & \textbf{95\% CI} & \textbf{Sig} \\ 
\hline
(Intercept)                        & 1.227 & 1.519 & -- \\
PR includes test                   & \cellcolor{mygreen3}1.244 & 1.054 & ** \\
Project age                        & \cellcolor{mygreen2}1.222 & 1.116 & *** \\
Cognitive complexity of PR          & \cellcolor{mygreen1}1.150 & 1.052 & ** \\
Contributor’s (+) emotion             & \cellcolor{mygreen0}1.054 & 1.007 & * \\
PR success rate of project          & \cellcolor{myred0}0.941 & 0.890 & * \\
\# of deleted files                    & \cellcolor{myred1}0.927 & 0.983 & * \\
\# of code smells                      & \cellcolor{myred2}0.871 & 0.791 & ** \\
\# of contributor’s previous PRs       & \cellcolor{myred3}0.780 & 0.712 & *** \\
\hline
\end{tabular}%
}
\vspace{1ex}
\textbf{Model fit:} $R^2_m = 0.027$, $R^2_c = 0.246$, ROC--AUC = 0.739
\end{table}

\textbf{RQ$_{5}$: For PRs that include a description, which factors are associated with whether a PR description includes specific elements?}

\begin{table}[htbp]
\centering
\footnotesize
\caption{Top significant positive and negative predictors for the inclusion of PR description elements}
\label{tab:summary_elements_predictors_singlecol}
\resizebox{\columnwidth}{!}{%
\begin{tabular}{p{0.7cm} p{3.2cm}p{3.5cm}}
\toprule
\textbf{Element} & \textbf{Top 3 Positive Predictors} & \textbf{Top 3 Negative Predictors} \\
\midrule
Purpose & 
1- Diff size (1.41) \newline
2- \#  of  commits (1.19) \newline
3- Changed files hotness (1.07) &
1- \# of contrib's previous PRs  (0.79) \newline
2- \#  of code smells (0.86) \newline
3- Contrib's daily workload (0.92) \\
\midrule
Code Exp & 
1- Cognitive complexity (1.17) \newline
2- Test size (1.07) \newline
3- Changed file hotness (1.06) &
1-\# of contrib’s previous PRs (0.87) \newline
2- \#  of code smells (0.92) \newline
3- Contrib's daily workload (0.95) \\
\midrule
Link &
1- PR includes test (1.29) \newline
2-  \#  of open PRs in project (1.24)\newline
3- Project age (1.22) &
1- Diff size (0.93)\newline
2- Contrib's daily workload  (0.97)\newline
3- \# of contrib’s previous PRs (0.99) \\
\midrule
Feed- back Type &
1- Diff size (1.26) \newline
2- \#  of commits (1.21) \newline
3- Contrib is core member (1.20) &
1- \# of contrib’s previous PRs (0.79) \newline
2- Contrib's daily workload (0.88) \newline
3- \#  of code smells (0.91) \\
\midrule
Order &
1- \#  of commits (1.54)\newline
2- diff size (1.53) &
1-Contrib's daily workload (0.45)\newline
2-Duplicated lines in code (0.82)
 \\
\midrule
Test Exp &
1- PR includes test  (1.72) \newline
2- Test size (1.41) \newline
3- Contrib's merge rate (1.23) &
1- \# of contrib’s previous PRs (0.74) \newline
2- Contrib's daily workload (0.87)\newline
3-  Project's PR merge rate (0.92)\\
\bottomrule
\end{tabular}}
\vspace{0.3em}
\begin{flushleft}
\tiny
\textbf{Purpose:} $R^2_m = 0.067$, $R^2_c = 0.1188$, AUC = 0.699 \quad
\textbf{Code Exp:} $R^2_m = 0.017$, $R^2_c = 0.124$, AUC = 0.657 \\
\textbf{Link:} $R^2_m = 0.074$, $R^2_c = 0.258$, AUC = 0.730 \quad
\textbf{Feedback Type:} $R^2_m = 0.059$, $R^2_c = 0.228$, AUC = 0.742 \\
\textbf{Order:} $R^2_m = 0.227$, $R^2_c = 0.441$, AUC = 0.908 \quad
\textbf{Test Exp:} $R^2_m = 0.102$, $R^2_c = 0.265$, AUC = 0.751
\end{flushleft}
\end{table}

Furthermore, we aimed to examine in what context PR descriptions are more likely to include specific elements. \autoref{tab:summary_elements_predictors_singlecol} summarizes the strongest predictors associated with the inclusion of each PR description element. The large gaps between marginal and conditional $R^2$ values for \textit{Purpose}, \textit{Code Explanation}, \textit{Link}, and \textit{Feedback Type} elements indicate that project-specific conventions substantially influence the presence of these elements. In contrast, \textit{Order} and \textit{Test Explanation} elements show more consistent behavior across projects. Across all six elements, several factors consistently emerge as significant. The most recurrent negative predictors are \textit{number of contributors’ previous PRs}, \textit{contributors’ daily workload}, and \textit{number of code smells in the changed files of the PR}. The most common positive predictors include \textit{diff size}, \textit{number of commits}, and \textit{test inclusion in the PR}. This pattern suggests that contributors tend to provide richer explanations when changes are bigger or related to testing. Conversely, the likelihood of including these elements decreases in PRs authored by more experienced or time-constrained contributors or PRs with lower code quality.

\section{Discussion}\label{sec:discussion}

Pull request descriptions provide a context for reviewers~\cite{comprehension}. Yet, it was still largely unknown what an effective PR description contains and whether the \textit{content} of these descriptions affects the code review process. Our study examined the value of PR descriptions through a mixed-methods approach that combined a gray literature review of guidelines, a large-scale analysis of software repositories, and a developer survey. Based on the gray literature review, we derived a taxonomy of eight PR description elements present in practitioner guidelines. We have extended an existing data set with further data from software repositories and analyzed whether description elements can be traced to a change in code review outcomes. Most of the analyzed relationships showed negligible importance. However, PRs that include in their description an explanation of the code and request a specific type of feedback are more likely to be merged.

Despite this limited evidence from recorded data, we found that PR descriptions are largely considered important by developers who reported that a missing PR description can hinder reviewers' efficiency in several ways: by missing context, lower understanding of the changes and the history of the software system. When we analyzed the software repositories, we could indeed observe that developers tend to include PR descriptions in older projects and PRs that are more complex. This suggests that, over time, teams develop a more mature documentation culture, carrying over the historical archive of changes to the software system in the PR description content. Developers also manifest awareness about the complexity of their changes, providing PR descriptions as a \textit{cognitive compensation mechanism}: when reasoning about a change becomes difficult, developers increase their communicative effort to reduce reviewer uncertainty. We have also found hindrances in providing the much-needed context to developers. Namely, heavy workloads or routine contributions reduce the likelihood of more detailed communication.

Similarly to PR descriptions being more valuable in certain contexts, the value of specific PR description elements varies across PRs. The \textbf{descriptive elements} such as the \textit{purpose}, \textit{code explanations}, \textit{link to related issues}, and \textit{test explanations} are present in approximately half of the analyzed PRs and considered important in most reviews by developers. These elements describe what was modified and why, thus retaining meaning even outside the review process. While code can often ``speak for itself,'' descriptive elements enrich its interpretation by making the rationale behind changes explicit. In doing so, they bridge the gap between implementation and intent, supporting both immediate review comprehension and long-term project understanding.

In contrast, the \textbf{interaction-oriented element}---especially, the \textit{type of feedback needed}---derives its value only within the collaborative context of code review. This element does not explain the change itself but instead helps coordinate the review interaction by directing reviewers' attention, clarifying expectations, and structuring the sequence of review activities. This element is included specially in large or multi-commit PRs and PRs submitted by the project's core members (see \autoref{tab:summary_elements_predictors_singlecol}). This connection explains our observation of software repositories where PR descriptions including this interaction element were linked to longer first response times, a higher number of review comments, and increased review latency. Even though the review process is slower and the communication more complex, inclusion of these interaction elements actually relates to a higher likelihood of the pull request being merged through promoting deeper engagement and more deliberate discussions among reviewers.

Recognizing the distinction between the roles of descriptive and interactive PR description elements underscores that PR descriptions serve as both descriptive and interaction-oriented functions. Therefore, both dimensions should be considered when designing contribution guidelines or developing automated tools that support the review process. Overall, these patterns indicate that PR documentation may not be a uniform habit but an adaptive behavior shaped by both organizational maturity and the complexity of the task at hand.

\subsection{Implications}

\subsubsection{For Practitioners}

Based on these findings, we identify actionable implications for practitioners, particularly software teams, project maintainers, and reviewers.

\noindent\textbf{Encourage Documentation Early.}  
Younger projects often lack established communication norms. Maintainers should introduce contribution guidelines that require or encourage descriptive PRs from the early stages of a project, ensuring that effective review practices develop before informal habits solidify.

\noindent\textbf{Context-Aware Guidance via Static Code Analysis.}  
Automated systems can apply these insights by using project- and code-level metrics (e.g., cognitive complexity, diff size, and presence of tests) to detect when additional explanation is beneficial. For example, tools could display adaptive prompts such as:\ ``This PR modifies complex or test-related code; please include a description.'' Such context-aware reminders can help maintain communication quality without overburdening contributors and ensure that critical context accompanies technically demanding or test-intensive changes.

\noindent\textbf{Specify the Feedback Type.}  
Encouraging contributors to explicitly state the kind of feedback they seek can make review interactions more focused and productive. Including a clear feedback request might lead reviewers' attention to the contributor's needs, thus reducing misunderstandings, facilitating richer discussions, and streamlining merge decisions. Teams can promote this practice by incorporating a prompt in PR templates or checklists, guiding contributors to specify one or more feedback themes.

\noindent\textbf{Support Reviewer Awareness.}
Reviewers can use similar cues to guide their actions. When encountering a large or complex PR without a description, reviewers may proactively request clarification, ensuring that key rationale and design decisions are not overlooked before merging.

\subsubsection{For Researchers}
The novel insights about PR Descriptions open avenues for future research. First, subsequent work can investigate how developers’ awareness of interaction-oriented elements evolves with experience and team culture, and how interventions, such as training, might influence these practices. Second, the design of intelligent PR assistants could move beyond static templates toward context-aware communication support that dynamically recommends missing elements based on project age, code characteristics, or review history. Our findings extend the understanding of developer communication in socio-technical systems by showing that PR description writing is not a routine documentation activity but a context-sensitive behavior. Developers adapt their communicative effort based on both technical characteristics (e.g., code complexity, test presence) and organizational factors (e.g., project age, contributor experience) to preserve knowledge about the changes to the software system and support reviewing effectiveness.

\section{Threats to Validity} \label{sec:threats-to-validity}

\noindent\textbf{Dataset recency and external validity.}
Our study relies on the publicly available GHTorrent release (2019). Since 2019, a significant shift in software development tooling through the widespread adoption of generative AI. This means that our repository-based findings may not fully reflect the most recent evolution of PR authoring practices. We cannot exclude the possibility that generative AI has influenced how PR descriptions are written in more recent years, but this potential effect is not measured in our dataset. Moreover, other changes in the practice of development since 2019 may have taken place, thus possibly changing our results if we considered data from a more recent timeframe. Nevertheless, our Gray Literature Review includes practitioner resources published both before and after 2019, and shows that the same PR description elements remain consistently recommended throughout the observed period, suggesting that the taxonomy studied in this paper remains relevant despite changes in tooling and practices.

\noindent\textbf{Measurement validity of first response time.}
The \emph{first response time} metric is obtained directly from the GHTorrent dataset. As a result, we did not apply additional filtering or bot identification specifically for this variable. Automated accounts can post early comments, which could bias this metric if bot-generated responses were present. To assess this risk, we manually inspected a random sample of 50 PR timelines used in our study and did not observe PRs in which the first responding account was a bot. However, we cannot exclude that bot-generated first responses may exist outside the inspected sample.

\noindent\textbf{Dependence on LLM-based Coding:}
We used Llama3.1 to identify PR description elements. Despite manual validation on a random subset of 100 PRs, the model’s interpretation may not perfectly align with human judgment. This introduces a potential source of misclassification bias in the presence or absence of elements.

\noindent\textbf{Model Specification and Control Selection.}
The mixed-effects regression models may be sensitive to the selection of control variables. Although we applied a systematic process to identify statistically significant controls (Section~\ref{sec:regression_models}), residual confounding cannot be fully ruled out. However, the consistency of results across both baseline and controlled models provides reassurance that the observed effects of PR description elements are robust.

\noindent\textbf{Survey Self-Selection Bias:}
The developer survey component of this study may suffer from self-selection bias. Participants who chose to respond might be more experienced with code review practices or have stronger opinions about PR descriptions than the general population of developers. This could skew the perceived importance of certain description elements.

\noindent\textbf{External Validity:}
We included only GitHub projects with a high proportion of external contributors and active pull-based workflows. While this ensured consistency and comparability, it may limit the generalizability of our findings. Projects with predominantly internal contributors, private repositories, or alternative review practices might exhibit different relationships between PR descriptions and review outcomes. Furthermore, other collaborative platforms may yield different behavioral patterns and effects. Hence, our conclusions are most applicable to open-source, pull-based development environments on GitHub. Additionally, project norms, organizational policies, and community culture can influence how contributors document and review code. These contextual factors may not generalize across ecosystems, especially in industrial or closed-source settings where review motivations differ.

\section{Conclusion}
\label{sec:conclusion}
Although practitioner guidelines frequently promote best practices for writing effective PR descriptions, there has been limited empirical evidence on which description content actually relates to review outcomes and how developers perceive the value of such information. In this paper, we presented a mixed-methods investigation of PR descriptions that connects community recommendations, large-scale repository evidence, and practitioner perceptions. First, through a Gray Literature Review, we consolidated fragmented guidance from practitioner and community sources into a taxonomy of eight PR description elements. Second, we implemented this taxonomy at scale and analyzed 80K PRs from 156 GitHub projects using mixed-effects regression models to examine how the presence of description elements relates to multiple review outcomes. Third, through a developer survey, we captured how practitioners perceive the importance of PR descriptions and how often specific elements are considered valuable during real-world code reviews. Finally, we investigated contextual factors that predict when PR descriptions are included and which elements are more likely to appear.

Our findings highlight that PR descriptions serve a dual role in practice. Descriptive elements (e.g., purpose, code explanation, testing information) are consistently perceived by developers as important for supporting comprehension of code changes. At the same time, our large-scale modeling shows that interaction-oriented elements---particularly specifying the type of feedback requested---are the most consistently associated with favorable review outcomes, such as likelihood of merge and reviewer engagement, despite appearing in fewer PRs. This indicates that PR descriptions should not be viewed only as explanations of code changes, but also as a way to shape reviewer attention, expectations, and interaction patterns during the review process. We further observe that description writing is context-sensitive rather than routine: PR descriptions are more common in mature projects and in complex code changes.

\subsection*{Acknowledgements}
 
S. Pirouzkhah and A. Bacchelli gratefully acknowledge the support of the Swiss National Science Foundation through the SNF Project No. 200021\_197227. The authors would also like to thank the Swiss Group for Original and Outside-the-box Software Engineering (CHOOSE) for their sponsoring.

\clearpage
\bibliographystyle{ACM-Reference-Format}
\bibliography{sample-base}

\end{document}